\documentclass[11pt]{article}
\usepackage[final]{acl}

\usepackage{times}
\usepackage{latexsym}
\usepackage[T1]{fontenc}
\usepackage[utf8]{inputenc}
\usepackage{microtype}
\usepackage{inconsolata}
\usepackage{graphicx}
\usepackage{amsmath,amssymb,booktabs,colortbl,multirow}
\usepackage{placeins}
\definecolor{panelgray}{RGB}{242,245,247}
\definecolor{TDeltaBetter}{RGB}{30,100,150}
\definecolor{TDeltaWorse}{RGB}{179,65,45}
\definecolor{TDeltaNeutral}{RGB}{85,85,85}
\newcommand{\res}{\mathcal{R}}
\newcommand{\univ}{\mathcal{U}}
\newcommand{\ind}{\mathbf{1}}
\title{Lost in the Fallback: How Reliably Do Typed Decision Models Use Fallback Options?}

\author{
  \textbf{Yuanbo Xie}\textsuperscript{1}\thanks{Equal contribution.},
  \textbf{Mingli Zhu}\textsuperscript{1}\footnotemark[1],
  \textbf{Tianyun Liu}\textsuperscript{2,3}
  \textbf{Yingjie Zhang}\textsuperscript{2,3},
  \textbf{Haixiao Xiong}\textsuperscript{1}\thanks{Corresponding authors.},
  \textbf{Tingwen Liu}\textsuperscript{2,3}\footnotemark[2] \\
  \textsuperscript{1}System Security Team, Li Auto Inc, China \\
  \textsuperscript{2}Institute of Information Engineering,
  Chinese Academy of Sciences, China \\
  \textsuperscript{3}School of Cyber Security,
  University of Chinese Academy of Sciences, China \\
  \texttt{\{xieyuanbo,zhumingli,xionghaixiao\}@lixiang.com} \\
  \texttt{\{liutianyun,zhangyingjie2023,liutingwen\}@iie.ac.cn}
}

\begin{document}
\maketitle
\begin{abstract}
Typed decision models (TDMs) are increasingly used in applications such as request routing and model selection, making reliable decisions critical to system performance. Fallback options accommodate categories outside regular choices, enabling deferral to human experts or more capable models. While recent works have examined adversarial vulnerabilities in TDMs, their ability to follow fallback rules in non-adversarial settings remains underexplored. 
We introduce the Paired Fallback Robustness (PFR) Test to evaluate two aspects of fallback reliability: whether models correctly switch between regular and fallback options, and whether they make consistent decisions under logically equivalent definitions of the same rule. 
We evaluate open-source decision models on three benchmarks, with controlled comparisons across architectures and parameter counts. 
We find that correct recognition does not reliably transfer to correct fallback options, and errors persist even when the correct category is explicitly supplied.
Moreover, logically equivalent formulations of the same fallback rule can also produce substantial disagreement in model decisions. 
These findings reveal a crucial reliability gap in current TDMs: fallback mechanisms designed to improve reliability can instead introduce errors. 
It highlights faithful execution of fallback rules and consistency under equivalent definitions as essential requirements for reliable typed decision-making.\footnote{We will make all code and data publicly available upon acceptance of the paper.}
\end{abstract}

\section{Introduction}
\begin{figure}[t]
\centering
\includegraphics[width=\columnwidth]{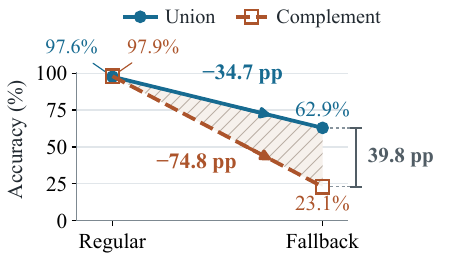}
\caption{Correct recognition does not ensure reliable fallback selection. On Intern/TREC50 inputs correctly classified in a separate task: (1) accuracy drops on fallback reassignment; (2) equivalent definitions yield different accuracy (hatched gap). Regular/Fallback denotes the required output. Union lists its members, while Complement excludes regular categories.}
\label{fig:teaser}
\end{figure}

Many software tasks, such as request routing \citep{ong2025routellm} and tool selection \citep{schick2023toolformer}, require choosing among predefined actions rather than generating free-form responses. Typed decision models (TDMs), such as TypeSafe's Jev \citep{typesafe2026systemone}, are designed to produce fast, type-safe decisions with associated probabilities and confidence estimates. 
However, in practical deployments, some requests fall outside the scope of predefined actions, while others require reasoning or expertise beyond the decision model's capabilities. Fallback options provide a mechanism for handling these cases by redirecting requests to alternative processing paths. 
For example, an address-change request can be routed to a dedicated service when available, or to general support through a fallback otherwise (Figure~\ref{fig:fallback-use-cases}a). Similarly, a fast System~1 controller can select Defer to invoke a more capable System~2 reasoning model \citep{chen2024frugalgpt} (Figure~\ref{fig:fallback-use-cases}b). 
 In both cases, the fallback provides an alternative handling path, but the model must correctly determine when to use it. 
Failing to select fallback can bypass necessary handling, whereas invoking it unnecessarily introduces additional cost or latency. 

Recent studies have examined adversarial vulnerabilities in TDMs, including adversarial attacks and prompt injection \citep{hu2026jevadvbench,wu2026decisionhijacking}. Concurrent work has investigated rejection behavior in arithmetic reasoning and candidate coverage \citep{zhong2026missing,luwu2026coverage}, as well as adherence to explicit option definitions \citep{sun2026typesafe}. Despite these advances, a fundamental question remains unexplored: 

\emph{How reliably do TDMs follow fallback rules under non-adversarial conditions?}

To answer this question, we introduce the \emph{Paired Fallback Robustness (PFR) Test}, which evaluates two complementary requirements of fallback reliability. \emph{Correct reassignment} examines whether models correctly switch from a regular option to fallback when the required assignment of the same input changes, testing their ability to faithfully execute fallback rules. \emph{Definition consistency} examines whether models preserve their decisions under logically equivalent formulations of the same fallback rule, testing whether decisions depend on the rule's meaning rather than its wording. Together, these paired comparisons assess whether models respond correctly to changes in the required assignment while remaining consistent when the underlying rule is unchanged.

Across three benchmarks, we find that \emph{even correctly recognized inputs can become lost during fallback reassignment}.
For example, on TREC50, Intern's accuracy drops from 97.9\% to 23.1\% when switching from regular to fallback assignment (Figure~\ref{fig:teaser}). These failures persist even when the correct category is explicitly supplied, including in matched comparisons across different model architectures. Moreover, logically equivalent formulations of the same fallback rule can lead to substantial disagreement in individual decisions, even when aggregate accuracy remains nearly unchanged. Together, these findings expose a critical reliability gap in current TDMs: fallback mechanisms intended to ensure appropriate handling can themselves become sources of error, even when the input category is correctly identified.

Our contributions are summarized as follows:
\begin{itemize}
    \item We identify and systematically study an underexplored reliability problem in TDMs: models may fail to follow fallback rules even when valid options and the necessary information are available, without adversarial inputs.
    \item We introduce the \emph{Paired Fallback Robustness (PFR) Test}, a systematic framework for evaluating fallback reliability
    through paired comparisons on the same inputs.
    \item Across diverse benchmarks, model architectures, and
    parameter scales, we show that correct category information
    and aggregate accuracy alone are insufficient to establish
    reliable fallback behavior.
\end{itemize}

\section{Preliminaries}
\label{sec:task}

\begin{figure*}[t]
\centering
\includegraphics[width=\textwidth]{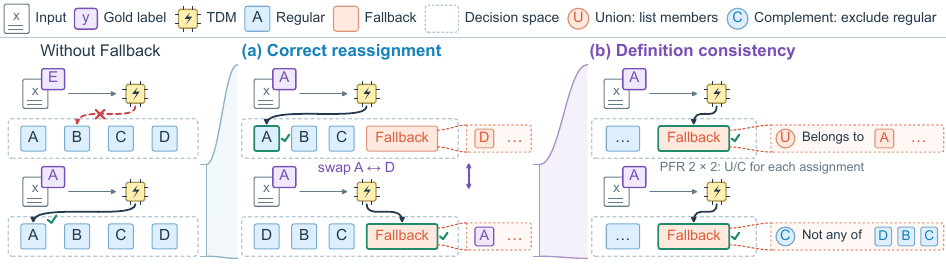}
\caption{PFR crosses two assignments with two equivalent definitions. Left: without fallback, inputs outside the regular categories have no valid option. (a) Moving the gold category into fallback changes the required output. (b) U/C preserve the required output within each assignment; gold-fallback is shown. Input, inventory and option count stay fixed. Left: missing coverage motivates fallback. Letters are illustrative categories; checks mark expected outputs and gold labels are evaluation annotations.}
\label{fig:pfr-protocol}
\end{figure*}

\paragraph{Typed choices.}
We study categorical Choice interfaces. A request contains an input $x$ and a finite option set $\mathcal{O}$ with natural-language selection criteria. The model returns an option $\hat{o}\in\mathcal{O}$ and associated scores; the application executes the corresponding action. Returning an allowed option ensures type validity, but decision correctness also requires satisfying its selection criterion.

\paragraph{Fallback options.}
A fallback option provides an alternative handling path when regular choices are unavailable or unsuitable. Its criterion is application-defined, for example, \textit{Other} may handle requests without a dedicated service, while \textit{Defer} may invoke a more capable model under a specified quality--cost policy.

\paragraph{Category-to-output assignment.}
We use category coverage to make fallback rules directly testable.  Let $\univ$ be the complete category inventory and $y\in\univ$ the input's true category. Categories in $S\subsetneq\univ$ have regular options, while fallback covers $\res=\univ\setminus S$. Representing each output by its category set, the required assignment is
\begin{equation}
g_S(y)=
\begin{cases}
\{y\}, & y\in S,\\
\res, & y\notin S.
\end{cases}
\label{eq:coverage}
\end{equation}
All category definitions are provided, and each input has exactly one valid destination. Gold labels are used to construct and score assignments but are disclosed to the model only in the category-information diagnostic. If $\hat g$ is the category set represented by $\hat{o}$, correctness means $\hat g=g_S(y)$. Changing $S$ can change the required output for the same input without changing its category. This tests fallback reassignment without introducing unknown categories or leaving inputs without valid destinations.

\paragraph{Equivalent definitions.}
We consider two logically equivalent ways to define fallback. \emph{Union} (U) explicitly lists the fallback categories in $\res$, while \emph{Complement} (C) defines fallback by excluding the regular categories $S$ from $\univ$. Both specify exactly the same category-to-output assignment. 

\section{Does Category Recognition Transfer to Fallback Selection?}
\label{sec:setup}
We test four 4B choice-selection configurations: Intern-Decision~\cite{internlm2026decision}, Kev~\cite{palmer2026kev4b}, JevK5~\cite{serikbay2026jevk5}, and SemIf~\cite{theoleecj2026semif}.
All receive identical requests with fifteen regular options and one fallback. Main evaluation uses class-balanced CLINC150 (450) \citep{larson-etal-2019-evaluation} and BANKING77 (385) \citep{casanueva-etal-2020-efficient} subsets, along with the full TREC50 test split (500) \citep{li-roth-2002-learning}. Unless specified otherwise, diagnostic and development panels each use 128 additional texts per dataset. The main sets measure performance across categories, while the diagnostic panels support repeated controlled comparisons on the same inputs. Sampling and model interfaces are introduced in Appendix~\ref{app:setup}.

\begin{table*}[t]
\centering
\begin{tabular*}{\textwidth}{@{\extracolsep{\fill}}c@{\,}|@{\,}l@{\,}|@{\,}c@{\,}c@{\,}c@{\,}|@{\,}c@{\,}c@{\,}c@{\,}|@{\,}c@{\,}c@{\,}c@{\,}|@{\,}c@{\,}c@{\,}c@{\,}|@{\,}c@{\,}c@{\,}|@{\,}c@{}}
\hline
\multirow{2}{*}{Dataset} & \multirow{2}{*}{Model} & \multicolumn{3}{@{\,}c@{\,}|@{\,}}{$A_+\uparrow$} & \multicolumn{3}{@{\,}c@{\,}|@{\,}}{$A_-\uparrow$} & \multicolumn{3}{@{\,}c@{\,}|@{\,}}{$F_+\downarrow$} & \multicolumn{3}{@{\,}c@{\,}|@{\,}}{$J\uparrow$} & \multicolumn{2}{@{\,}c@{\,}|@{\,}}{DR$\downarrow$} & \multirow{2}{*}{$P_4\uparrow$}\\
\cline{3-16}
& & U & C & $\Delta$ & U & C & $\Delta$ & U & C & $\Delta$ & U & C & $\Delta$ & R & F & \\
\hline
 & Intern & 91.1 & 90.7 & \textcolor{TDeltaWorse}{$-0.4$} & 68.7 & 75.8 & \textcolor{TDeltaBetter}{$+7.1$} & 6.9 & 7.6 & \textcolor{TDeltaWorse}{$+0.7$} & 62.2 & 68.4 & \textcolor{TDeltaBetter}{$+6.2$} & 2.7 & 28.7 & 52.0\\
 & Kev & 86.4 & 88.4 & \textcolor{TDeltaBetter}{$+2.0$} & 94.2 & 92.9 & \textcolor{TDeltaWorse}{$-1.3$} & 11.3 & 9.8 & \textcolor{TDeltaBetter}{$-1.6$} & 82.9 & 83.1 & \textcolor{TDeltaBetter}{$+0.2$} & 3.6 & 3.8 & 80.2\\
 & SemIf & 78.2 & 78.0 & \textcolor{TDeltaWorse}{$-0.2$} & 78.7 & 93.3 & \textcolor{TDeltaBetter}{$\mathbf{+14.7}$} & 15.8 & 20.4 & \textcolor{TDeltaWorse}{$\mathbf{+4.7}$} & 62.4 & 72.9 & \textcolor{TDeltaBetter}{$+10.4$} & 9.3 & 20.2 & 58.2\\
\multirow{-4}{*}{\small\shortstack{CLINC150\\$n=450$}} & JevK5 & 93.8 & 92.7 & \textcolor{TDeltaWorse}{$-1.1$} & 57.3 & 59.1 & \textcolor{TDeltaBetter}{$+1.8$} & 1.6 & 3.6 & \textcolor{TDeltaWorse}{$+2.0$} & 55.8 & 55.8 & \textcolor{TDeltaNeutral}{$+0.0$} & 3.3 & 33.3 & 40.9\\
\hline
 & Intern & 81.3 & 82.1 & \textcolor{TDeltaBetter}{$+0.8$} & 74.0 & 67.5 & \textcolor{TDeltaWorse}{$-6.5$} & 11.2 & 10.4 & \textcolor{TDeltaBetter}{$-0.8$} & 62.9 & 57.4 & \textcolor{TDeltaWorse}{$-5.5$} & 3.4 & 21.0 & 50.6\\
 & Kev & 89.4 & 83.1 & \textcolor{TDeltaWorse}{$-6.2$} & 96.4 & 97.7 & \textcolor{TDeltaBetter}{$+1.3$} & 8.8 & 15.6 & \textcolor{TDeltaWorse}{$\mathbf{+6.8}$} & 87.5 & 82.1 & \textcolor{TDeltaWorse}{$-5.5$} & 7.3 & 1.8 & 81.0\\
 & SemIf & 69.6 & 73.0 & \textcolor{TDeltaBetter}{$+3.4$} & 89.4 & 87.8 & \textcolor{TDeltaWorse}{$-1.6$} & 25.7 & 23.4 & \textcolor{TDeltaBetter}{$-2.3$} & 63.9 & 65.2 & \textcolor{TDeltaBetter}{$+1.3$} & 7.8 & 10.9 & 57.7\\
\multirow{-4}{*}{\small\shortstack{BANKING77\\$n=385$}} & JevK5 & 85.7 & 86.2 & \textcolor{TDeltaBetter}{$+0.5$} & 53.5 & 47.3 & \textcolor{TDeltaWorse}{$-6.2$} & 3.6 & 6.0 & \textcolor{TDeltaWorse}{$+2.3$} & 49.9 & 41.8 & \textcolor{TDeltaWorse}{$-8.1$} & 7.5 & 28.1 & 33.5\\
\hline
 & Intern & 80.4 & 83.8 & \textcolor{TDeltaBetter}{$+3.4$} & 59.4 & 22.6 & \textcolor{TDeltaWorse}{$\mathbf{-36.8}$} & 7.2 & 1.4 & \textcolor{TDeltaBetter}{$\mathbf{-5.8}$} & 51.8 & 21.0 & \textcolor{TDeltaWorse}{$-30.8$} & 6.8 & 39.8 & 18.2\\
 & Kev & 84.2 & 87.6 & \textcolor{TDeltaBetter}{$+3.4$} & 88.0 & 74.0 & \textcolor{TDeltaWorse}{$\mathbf{-14.0}$} & 9.6 & 6.0 & \textcolor{TDeltaBetter}{$\mathbf{-3.6}$} & 78.0 & 68.2 & \textcolor{TDeltaWorse}{$-9.8$} & 5.8 & 15.8 & 64.0\\
 & SemIf & 82.0 & 83.4 & \textcolor{TDeltaBetter}{$+1.4$} & 72.8 & 60.4 & \textcolor{TDeltaWorse}{$\mathbf{-12.4}$} & 9.2 & 8.2 & \textcolor{TDeltaBetter}{$-1.0$} & 63.6 & 52.0 & \textcolor{TDeltaWorse}{$-11.6$} & 8.2 & 24.0 & 45.6\\
\multirow{-4}{*}{\small\shortstack{TREC50\\$n=500$}} & JevK5 & 82.8 & 85.4 & \textcolor{TDeltaBetter}{$+2.6$} & 57.8 & 24.2 & \textcolor{TDeltaWorse}{$\mathbf{-33.6}$} & 4.0 & 1.4 & \textcolor{TDeltaBetter}{$\mathbf{-2.6}$} & 53.2 & 22.8 & \textcolor{TDeltaWorse}{$-30.4$} & 6.4 & 37.6 & 20.8\\
\hline
\end{tabular*}
\caption{Main PFR results (\%) under direct selection with fallback named Other. U/C: union/complement; $\Delta=C-U$ (pp). $A_+$: regular accuracy; $A_-$: fallback recall; $F_+$: false fallback rate; $J$: joint correctness under both assignments. DR: U/C disagreement under regular (R) or fallback (F) assignment. $P_4$: correct on all four requests. Blue/orange: improvement/deterioration. Bold $A_-$/$F_+$ deltas: Holm-adjusted $p<.05$; paired intervals: Figure~\ref{fig:primary}.}
\label{tab:primary-absolute}
\label{tab:paired}
\end{table*}

\subsection{Paired Fallback Robustness Test}
PFR constructs four requests per input by combining two assignments with two logically equivalent definitions (Figure~\ref{fig:pfr-protocol}). The input, full inventory and option count remain fixed. 
This design tests whether the model changes its decision when the required assignment changes and preserves its decision when the assignment remains the same.

\paragraph{Requirement 1: correct reassignment.}
\emph{Follow the category's assigned destination.}
For each input, fifteen non-gold categories are sampled independently of model predictions. Fourteen remain regular choices in both conditions. The \textbf{gold-regular} condition adds $y$; \textbf{gold-fallback} replaces it with the remaining sampled category in the same slot. Every model receives the same pair. 
The required output changes from the regular option for $y$ to fallback, while the number and positions of options remain fixed.

\paragraph{Requirement 2: definition consistency.}
\emph{Preserve the decision under equivalent descriptions.}
Within each assignment, U lists the fallback members and C excludes regular categories. Both definitions specify the same category-to-output assignment, so the required output is unchanged.

\paragraph{Measuring both requirements.}
For $n$ texts, superscripts $+$/$-$ denote gold-regular/gold-fallback, with predictions $\hat g_i^\pm$ and fallback sets $\res_i^\pm$. We measure:
\begin{align}
A_+&=n^{-1}\sum_i\ind[\hat g_i^+=\{y_i\}],\\
F_+&=n^{-1}\sum_i\ind[\hat g_i^+=\res_i^+],\\
A_-&=n^{-1}\sum_i\ind[\hat g_i^-=\res_i^-].
\end{align}
Here $A_+$ is regular accuracy, $A_-$ fallback recall, and $F_+$ false fallback rate. Since incorrect regular choices also reduce $A_+$, we have $F_+ + A_+ \leq 1$.

\emph{Joint correctness} ($J$) is the fraction of inputs correctly handled under both assignments for a given definition. \emph{Disagreement rate} (DR) measures the fraction of inputs receiving different decisions under U and C within the same assignment, including cases where both decisions are incorrect. 
Thus $J$ measures successful reassignment, while DR tests definition consistency. 
The \emph{PFR pass rate} ($P_4$) is the fraction of inputs correctly handled in all four requests, jointly satisfying both requirements.
These metrics capture different aspects of reliability. By construction, $P_4\leq\min(J_U,J_C)$, but equal $J$ values do not imply that the same inputs are correctly handled under both definitions. 
Similarly, net score changes $\Delta=C-U$ can hide switches in opposite directions. Intervals use 95\% paired text-bootstrap resampling; primary U/C contrasts use Holm-corrected McNemar tests, and diagnostic intervals are descriptive (Appendix~\ref{app:statistics}).

Joint evaluation also avoids trivial behavior: always selecting fallback preserves agreement but fails regular cases; avoiding fallback removes false fallback while missing required transfers. PFR therefore tests responsiveness to changed assignments and invariance to equivalent descriptions on the same inputs.

\subsection{Correct Classifications Can Be Lost on Reassignment}
\label{sec:recognition}
Under direct selection, $P_4$ ranges from 18.2\% to 81.0\% across the twelve model--dataset settings (Table~\ref{tab:primary-absolute}). To examine whether these failures also occur on inputs that models can correctly classify, we use a separate full-inventory classification task as a reference. Native option limits allow this evaluation for Kev on all three datasets and Intern on TREC. We then evaluate reference-correct inputs under the PFR assignments without providing their category labels to the model.

Intern correctly classifies 329 of the 500 TREC inputs in the reference task. On these same inputs, accuracy under C drops from 97.9\% in the gold-regular condition to 23.1\% in the gold-fallback condition (Figure~\ref{fig:teaser}). Thus, correct classification in the reference task does not guarantee correct fallback selection. Paired examples in Appendix~\ref{app:paired-examples} illustrate failures under both definitions.

\subsection{Supplying the Correct Category Is Not Sufficient}
\label{sec:known-category}
We next provide the correct category directly to the model. On TREC, supplying the fine-category label makes every JevK5 gold-regular decision correct under both U and C, yet fallback recall reaches only 71.1\% and 14.8\%, respectively. When the input text is removed and only the correct label is provided, C recall increases to 50.0\%, while regular accuracy remains perfect (Figure~\ref{fig:information}). These results show that fallback errors persist even when the model does not need to infer the category from the input.

\begin{figure}[t]
\centering
\includegraphics[width=\columnwidth]{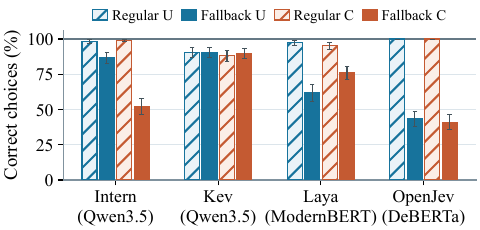}
\caption{Providing the correct category does not eliminate fallback errors. The matched TREC task uses six categories and four options. Bars show regular and fallback accuracy under Union (U) and Complement (C), with paired text-bootstrap intervals.}
\label{fig:backbones}
\end{figure}

\paragraph{Assignment across backbones.}
We further compare Intern and Kev with Laya and OpenJev \citep{convai2026laya,kotoba2026openjev} on a matched task containing 256 TREC inputs, six coarse categories, and four options. Even with the correct category provided, OpenJev achieves perfect regular accuracy but only 43.4\% and 41.0\% fallback recall under U and C, respectively (Figure~\ref{fig:backbones}). However, these failures are not universal. In a separate sixteen-option category-only task, Kev achieves 99.2\% and 98.4\% joint correctness on BANKING77 under U and C, respectively.

\paragraph{Takeaway.}
\emph{Correct category recognition does not guarantee correct fallback assignment. Errors can persist even when the correct category is explicitly provided.}

\section{Do Equivalent Definitions Preserve Decisions?}
We examine whether logically equivalent fallback definitions lead to consistent decisions. Using the PFR setup from Section~\ref{sec:setup}, we compare Union (U) and Complement (C) on the same inputs and category assignments, keeping the option set and category membership fixed. We first examine differences in accuracy and individual decisions, then test their sensitivity to model size and presentation, and finally extend the analysis to Fast/Defer policies.

\label{sec:results}
\subsection{Similar Performance, Different Decisions}
Similar aggregate scores can conceal different decisions. JevK5's CLINC fallback recalls differ by only 1.8\% between U and C, yet one-third of its choices change. Although $J$ is 55.8\% under both definitions, only 40.9\% of inputs pass the full PFR Test. Gold-fallback disagreement exceeds 10\% in ten of the twelve model--dataset pairs (Table~\ref{tab:primary-absolute}).

The direction of the accuracy change depends on the task (Figure~\ref{fig:primary}). C lowers TREC recall for all four configurations after correction, but increases both recall and false fallback for SemIf on CLINC. On BANKING, no recall contrast survives correction. Equivalent descriptions can therefore change both which inputs reach fallback and how often that decision is correct; neither description is uniformly preferable.

\subsection{Higher Recall Does Not Ensure Robustness}
\label{sec:sensitivity}
The discrepancy persists across model sizes and controlled changes in presentation. Each comparison holds the texts fixed.

\begin{figure*}[t]
\centering
\includegraphics[width=\textwidth]{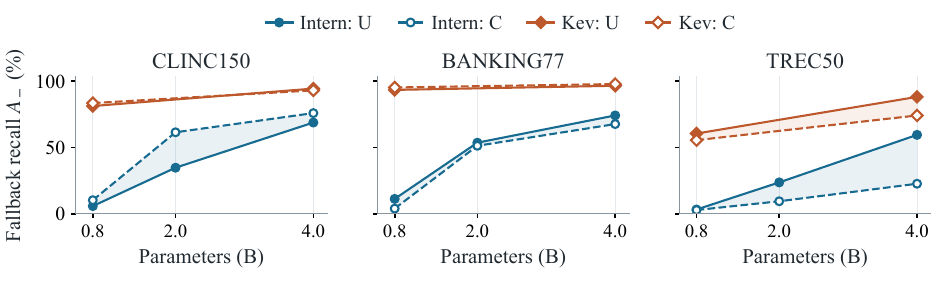}
\caption{Higher fallback recall does not necessarily reduce the U/C gap. Each model family uses the same evaluation texts and assignments across parameter counts. Solid/dashed lines denote U/C; colors distinguish Intern and Kev.}
\label{fig:scale-extension}
\end{figure*}

\paragraph{Higher recall does not ensure a smaller definition gap.}
Across Intern 0.8B/2B/4B and Kev 0.8B/4B, fallback recall increases under both definitions without consistently narrowing their difference (Figure~\ref{fig:scale-extension}). Intern's TREC gap grows from 0.4 to 36.8 points between 0.8B and 4B; CLINC's largest gap instead occurs at 2B. At 0.8B, both Intern/TREC recalls are below 4\%: the small gap coexists with almost complete fallback failure.

\begin{figure}[t]
\centering
\includegraphics[width=\columnwidth]{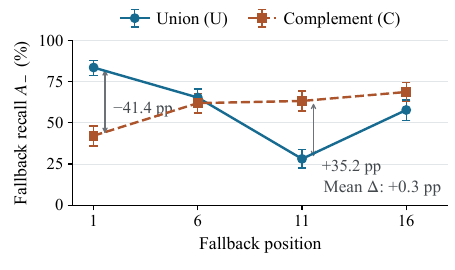}
\caption{Position reverses the U/C gap for JevK5/CLINC150. Points average three choice sets with paired-text intervals. Averaging across all four positions nearly cancels the difference, $\Delta=C-U$.}
\label{fig:position-interaction}
\end{figure}

\paragraph{Position can reverse the definition gap.}
For JevK5 on CLINC150, the C-minus-U recall gap changes from $-41.4$ points at fallback position 1 to $+35.2$ at position 11, while mean regular accuracy remains 94.8--97.1\%. Averaging across four positions and three choice sets leaves a gap of just $+0.3$ points (Figure~\ref{fig:position-interaction}). A small average can thus conceal large definition effects of opposite sign.

\paragraph{The gap survives matched controls.}
\label{sec:matched}
Matching token counts and category mentions still leaves TREC recall losses of 3.5--45.7 points under negative wording (Figure~\ref{fig:matched}). The direction also persists across three wording pairs and same-parent or different-parent replacement categories. Splitting an unchanged union list over lines lowers recall as well. These controls show that the observed sensitivity extends across several presentations of the same assignment (Appendix~\ref{app:robustness}).

\subsection{From Category Coverage to Defer Policies}
\label{sec:routing}
Does the failure extend to deciding when to delegate? On TREC, we supply hypothetical quality estimates $q_{\mathrm{fast}},q_{\mathrm{slow}}$ and extra cost $c$. Defer is required iff $q_{\mathrm{slow}}-\lambda c>q_{\mathrm{fast}}$; ties select Fast. Changing only $\lambda$ switches the required action. Each policy is expressed through a gain inequality or an equivalent utility-complement criterion. Here $J$ requires correctness under both policies.

Joint policy correctness is zero for all four models under both criteria. SemIf and JevK5 always choose Fast: perfect agreement coexists with zero $J$. Adding precomputed utilities or the correct Defer predicate with its action mapping changes information content, length and framing.

\begin{figure}[t]
\centering
\includegraphics[width=\columnwidth]{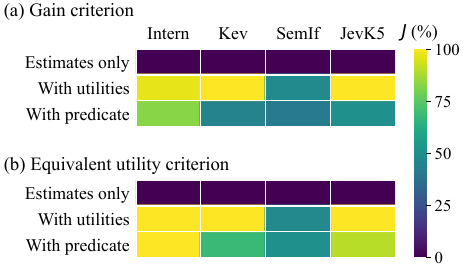}
\caption{Policy execution varies with the supplied information. $J$ requires correct Fast and Defer choices under the two policy weights. Utilities and the correct predicate are separate additions to the estimates; panels use equivalent gain (a) and utility (b) criteria.}
\label{fig:routing}
\end{figure}

Precomputed utilities make Kev and JevK5 correct on every paired request, while SemIf remains below 50\% joint correctness under both criteria. The correct predicate leaves errors for every model (Figure~\ref{fig:routing}). A category-based policy likewise changes decisions under equivalent rules across all three datasets (Appendix~\ref{app:routing-protocol}).

Errors also occur in the opposite direction. With zero quality gain and $\lambda=1$, Intern correctly selects Fast on all 64 inputs under gain wording but unnecessarily defers on 32 under the equivalent utility-complement criterion (Appendix~\ref{app:defer-policy}).

\paragraph{Takeaway.}
\emph{Reliable fallback use requires both the right change of output and invariance to equivalent rules. Neither similar accuracy nor agreement alone establishes both.}

\section{What Does Reliable Fallback Selection Require?}
\label{sec:membership}
We investigate this gap through three comparisons: deterministic mapping, explicit category membership, and selection rules applied to fixed scores. All use the main category-coverage sets.

\subsection{Separating Recognition from Assignment}
\label{sec:programmatic}
As a diagnostic, we use a classify-then-map reference where full-category prediction is supported. The model first predicts the input category, which is then deterministically mapped to a regular or fallback option using $g_S$. By construction, decisions remain consistent under U and C, while joint correctness $J$ equals the underlying classification accuracy. For Intern on TREC, this approach achieves $J=65.8\%$, compared with 21.0\% under direct C selection. However, classification errors still produce 22.8\% false fallback.

\begin{table*}[t]
\centering
{\small
\renewcommand{\arraystretch}{1.2}
\begin{tabular*}{\textwidth}{@{\extracolsep{\fill}}c@{\,}|@{\,}l@{\,}|@{\,}c@{\,}c@{\,}c@{\,}|@{\,}c@{\,}c@{\,}c@{\,}|@{\,}c@{\,}c@{\,}c@{}}
\hline
\multirow{2}{*}{{\small\shortstack{Dataset}}} & \multirow{2}{*}{Model} & \multicolumn{3}{@{\;}c@{\;}|@{\;}}{Direct} & \multicolumn{3}{@{\;}c@{\;}|@{\;}}{Membership} & \multicolumn{3}{@{\;}c@{}}{Threshold}\\
\cline{3-11}
 & & $J_U\uparrow$ & $J_C\uparrow$ & DR-F$\downarrow$ & $J_U\uparrow$ & $J_C\uparrow$ & DR-F$\downarrow$ & $J_U\uparrow$ & $J_C\uparrow$ & DR-F$\downarrow$\\
\hline
\multirow{4}{*}{{\small\shortstack{CLINC150}}} & Intern & 62.2 & 68.4 & 28.7 & 79.6\,{\small\textcolor{TDeltaBetter}{$(+17.3)$}} & 79.6\,{\small\textcolor{TDeltaBetter}{$(+11.1)$}} & 6.4\,{\small\textcolor{TDeltaBetter}{$(-22.2)$}} & 73.6\,{\small\textcolor{TDeltaBetter}{$(+11.3)$}} & 74.9\,{\small\textcolor{TDeltaBetter}{$(+6.4)$}} & 11.6\,{\small\textcolor{TDeltaBetter}{$(-17.1)$}}\\
 & Kev & 82.9 & 83.1 & 3.8 & 84.7\,{\small\textcolor{TDeltaBetter}{$(+1.8)$}} & 85.1\,{\small\textcolor{TDeltaBetter}{$(+2.0)$}} & 0.9\,{\small\textcolor{TDeltaBetter}{$(-2.9)$}} & 84.4\,{\small\textcolor{TDeltaBetter}{$(+1.6)$}} & 81.6\,{\small\textcolor{TDeltaWorse}{$(-1.6)$}} & 4.2\,{\small\textcolor{TDeltaWorse}{$(+0.4)$}}\\
 & SemIf & 62.4 & 72.9 & 20.2 & 72.7\,{\small\textcolor{TDeltaBetter}{$(+10.2)$}} & 71.8\,{\small\textcolor{TDeltaWorse}{$(-1.1)$}} & 1.3\,{\small\textcolor{TDeltaBetter}{$(-18.9)$}} & 64.7\,{\small\textcolor{TDeltaBetter}{$(+2.2)$}} & 72.0\,{\small\textcolor{TDeltaWorse}{$(-0.9)$}} & 15.6\,{\small\textcolor{TDeltaBetter}{$(-4.7)$}}\\
 & JevK5 & 55.8 & 55.8 & 33.3 & 80.4\,{\small\textcolor{TDeltaBetter}{$(+24.7)$}} & 82.0\,{\small\textcolor{TDeltaBetter}{$(+26.2)$}} & 2.9\,{\small\textcolor{TDeltaBetter}{$(-30.4)$}} & 82.0\,{\small\textcolor{TDeltaBetter}{$(+26.2)$}} & 72.2\,{\small\textcolor{TDeltaBetter}{$(+16.4)$}} & 10.7\,{\small\textcolor{TDeltaBetter}{$(-22.7)$}}\\
\hline
\multirow{4}{*}{{\small\shortstack{BANKING77}}} & Intern & 62.9 & 57.4 & 21.0 & 73.5\,{\small\textcolor{TDeltaBetter}{$(+10.6)$}} & 70.6\,{\small\textcolor{TDeltaBetter}{$(+13.2)$}} & 7.0\,{\small\textcolor{TDeltaBetter}{$(-14.0)$}} & 69.9\,{\small\textcolor{TDeltaBetter}{$(+7.0)$}} & 67.5\,{\small\textcolor{TDeltaBetter}{$(+10.1)$}} & 5.7\,{\small\textcolor{TDeltaBetter}{$(-15.3)$}}\\
 & Kev & 87.5 & 82.1 & 1.8 & 80.8\,{\small\textcolor{TDeltaWorse}{$(-6.8)$}} & 80.5\,{\small\textcolor{TDeltaWorse}{$(-1.6)$}} & 0.0\,{\small\textcolor{TDeltaBetter}{$(-1.8)$}} & 88.3\,{\small\textcolor{TDeltaBetter}{$(+0.8)$}} & 82.1\,{\small\textcolor{TDeltaNeutral}{$(0.0)$}} & 1.8\,{\small\textcolor{TDeltaNeutral}{$(0.0)$}}\\
 & SemIf & 63.9 & 65.2 & 10.9 & 61.6\,{\small\textcolor{TDeltaWorse}{$(-2.3)$}} & 60.8\,{\small\textcolor{TDeltaWorse}{$(-4.4)$}} & 2.1\,{\small\textcolor{TDeltaBetter}{$(-8.8)$}} & 62.6\,{\small\textcolor{TDeltaWorse}{$(-1.3)$}} & 62.6\,{\small\textcolor{TDeltaWorse}{$(-2.6)$}} & 2.9\,{\small\textcolor{TDeltaBetter}{$(-8.1)$}}\\
 & JevK5 & 49.9 & 41.8 & 28.1 & 69.9\,{\small\textcolor{TDeltaBetter}{$(+20.0)$}} & 69.6\,{\small\textcolor{TDeltaBetter}{$(+27.8)$}} & 7.5\,{\small\textcolor{TDeltaBetter}{$(-20.5)$}} & 69.6\,{\small\textcolor{TDeltaBetter}{$(+19.7)$}} & 59.7\,{\small\textcolor{TDeltaBetter}{$(+17.9)$}} & 10.9\,{\small\textcolor{TDeltaBetter}{$(-17.1)$}}\\
\hline
\multirow{4}{*}{{\small\shortstack{TREC50}}} & Intern & 51.8 & 21.0 & 39.8 & 51.2\,{\small\textcolor{TDeltaWorse}{$(-0.6)$}} & 29.0\,{\small\textcolor{TDeltaBetter}{$(+8.0)$}} & 28.6\,{\small\textcolor{TDeltaBetter}{$(-11.2)$}} & 51.2\,{\small\textcolor{TDeltaWorse}{$(-0.6)$}} & 53.4\,{\small\textcolor{TDeltaBetter}{$(+32.4)$}} & 17.0\,{\small\textcolor{TDeltaBetter}{$(-22.8)$}}\\
 & Kev & 78.0 & 68.2 & 15.8 & 72.4\,{\small\textcolor{TDeltaWorse}{$(-5.6)$}} & 71.0\,{\small\textcolor{TDeltaBetter}{$(+2.8)$}} & 4.2\,{\small\textcolor{TDeltaBetter}{$(-11.6)$}} & 77.4\,{\small\textcolor{TDeltaWorse}{$(-0.6)$}} & 70.2\,{\small\textcolor{TDeltaBetter}{$(+2.0)$}} & 13.8\,{\small\textcolor{TDeltaBetter}{$(-2.0)$}}\\
 & SemIf & 63.6 & 52.0 & 24.0 & 60.2\,{\small\textcolor{TDeltaWorse}{$(-3.4)$}} & 58.4\,{\small\textcolor{TDeltaBetter}{$(+6.4)$}} & 16.2\,{\small\textcolor{TDeltaBetter}{$(-7.8)$}} & 68.4\,{\small\textcolor{TDeltaBetter}{$(+4.8)$}} & 64.0\,{\small\textcolor{TDeltaBetter}{$(+12.0)$}} & 15.8\,{\small\textcolor{TDeltaBetter}{$(-8.2)$}}\\
 & JevK5 & 53.2 & 22.8 & 37.6 & 60.6\,{\small\textcolor{TDeltaBetter}{$(+7.4)$}} & 45.8\,{\small\textcolor{TDeltaBetter}{$(+23.0)$}} & 23.4\,{\small\textcolor{TDeltaBetter}{$(-14.2)$}} & 62.6\,{\small\textcolor{TDeltaBetter}{$(+9.4)$}} & 58.8\,{\small\textcolor{TDeltaBetter}{$(+36.0)$}} & 13.4\,{\small\textcolor{TDeltaBetter}{$(-24.2)$}}\\
\hline
\end{tabular*}
}
\caption{Correctness and consistency respond differently to intervention. $J_U,J_C$: joint correctness under U/C; DR-F: gold-fallback U/C disagreement (\%). Parentheses show changes from Direct (pp); blue/orange indicate improvement/deterioration. Evaluation sets match Table~\ref{tab:primary-absolute}.}
\label{tab:intervention-consistency}
\end{table*}

\subsection{Making the Assignment Explicit}
Can the model directly use the same mapping? A membership table labels regular categories as T1 and fallback categories as T2, without revealing the category of the input. Under U, fallback is defined by T2; under C, it is defined by excluding T1. Moving a category into fallback changes its entry from T1 to T2. This intervention makes the assignment explicit while changing its presentation. Table~\ref{tab:intervention-consistency} separates gains in joint correctness from changes in disagreement: explicit membership need not improve both together. With explicit membership, SemIf's fallback recall on CLINC reaches at least 98\% under both definitions, while gold-fallback disagreement drops from 20.2\% to 1.3\%. For Intern on TREC under C, $J$ increases from 21.0\% to 29.0\%.

\subsection{Limits of a Shared Selection Rule}
We next hold the requests and scores fixed. A threshold selects fallback when its log-score exceeds the largest regular log-score by more than $\tau$; otherwise the best regular option is selected. For each model--dataset setting, we select a threshold that maximizes assignment-balanced accuracy $B=(A_++A_-)/2$, averaged over U and C, on a separate development set. The threshold is then fixed for testing. Changing the selection rule recovers decisions that direct selection misses. Intern's TREC complement recall rises from 22.6\% to 75.6\%, but false fallback rises from 1.4\% to 20.6\% (Figure~\ref{fig:threshold-example}). The scores therefore support substantial recovery at the cost of unnecessary fallback selections.

\paragraph{What can a shared threshold recover?}
A threshold preserves the ranking of regular options. Passing PFR requires a correct regular winner and compatible thresholds across both assignments and definitions. Individual recoverability does not guarantee recovery under a shared threshold. For Intern/TREC, individual thresholds can satisfy all four PFR requests for 359 of 500 inputs, compared with at most 205 under a single shared threshold and 200 under the development-selected threshold. This gap appears across all settings (Figure~\ref{fig:recovery-limits}), as individually recoverable inputs may require incompatible thresholds. The individual and shared oracles use test labels to maximize four-request correctness at fixed scores, whereas the development rule optimizes mean $B$. Further derivations appear in Appendix~\ref{app:recovery-limits}.

\subsection{Agreement and Low False Fallback Are Insufficient}
Greater agreement does not establish correctness: U and C can consistently select the same wrong option. Membership information reduces gold-fallback disagreement in all twelve model--dataset pairs, yet Kev/BANKING becomes perfectly consistent on those inputs while its joint correctness declines under both definitions (Table~\ref{tab:intervention-consistency}). Membership information and thresholding can correct some errors while introducing others; thresholding may also improve fallback agreement at the expense of regular agreement (Figure~\ref{fig:repair-overlap}).

Low false fallback answers a different question. $F_+$ measures unnecessary fallback on gold-regular inputs; it leaves required transfers on gold-fallback inputs untested. On the TREC naming panel, changing SemIf's fallback name from \textit{Other} to \textit{None} lowers $F_+$ from 10.2\% to 6.3\%, while joint correctness falls from 43.8\% to 22.7\% (Appendix~\ref{app:fallback-names}). An order-statistic rule fitted on the development panels targets $F_+$ at 5\% \citep{angelopoulos2024risk}. Using the larger of the two thresholds calibrated to control $F_+$ under U and C, nine of twelve settings meet this empirical target under both definitions, yet fallback disagreement ranges from 10.7\% to 44.9\% and four-request correctness can be as low as 3.6\% (Figure~\ref{fig:risk-pfr}).

\paragraph{Implications for decision models.}
Evaluation should track each input's outcomes across both category assignments and equivalent descriptions. This distinguishes corrected decisions from newly introduced errors, and agreement on correct outputs from agreement on incorrect ones.

\paragraph{Takeaway.}
\emph{Assess correctness and consistency together: fewer unnecessary fallbacks or more agreement can still conceal failed transfers and newly introduced errors.}

\section{Related Work}

\paragraph{Language models and TDMs.}
Language models support software decisions through tool use \citep{schick2023toolformer} and structured generation \citep{geng-etal-2023-grammar}. Jev exposes typed choices \citep{typesafe2026systemone}; LLM-as-Jev studies next-token decision readouts \citep{li2026llm2jev}. \citet{zhong2026missing} pair available and missing arithmetic answers, while \citet{luwu2026coverage} study rejection-policy transfer and different sources of missing candidates. PFR holds the full category inventory explicit and changes its output partition. It tests whether recognized categories reach their assigned output by jointly evaluating two assignments and two equivalent definitions on each input.

\paragraph{Abstention and learned delegation.}
Fallback may implement abstention \citep{gangrade2021selective}, clarification \citep{saeidi2018sharc,hengst-etal-2024-conformal}, or delegation. SelectiveNet jointly learns prediction and rejection for a target coverage \citep{geifman2019selectivenet}; learning to defer learns a classifier and rejector from expert decisions \citep{mozannar2020defer}. RouteLLM learns model selection from preferences \citep{ong2025routellm}, while reinforcement-learning routers optimize delegation or stopping from downstream rewards \citep{yu2026salmonnduo,huang2026rlcascade}. These methods learn which action is useful. PFR tests execution of a supplied assignment or policy, including when the relevant category or decision information is already given.

\paragraph{Robustness, consistency and calibration.}
CheckList tests expected changes and invariances \citep{ribeiro-etal-2020-beyond}; order and prompt studies expose selection biases \citep{pmlr-v139-zhao21c,pezeshkpour-hruschka-2024-large,wei-etal-2024-unveiling,salinas2024butterfly}. \citet{sun2026typesafe} rebind option names to fixed definitions and score the selected meaning. PFR changes category membership, then restates each assignment equivalently. Jointly scoring the resulting requests reveals losses concealed by equal marginal accuracy. Grouped prediction \citep{goren2024hierarchical,jain2023amendment}, probabilistic coherence \citep{chenli2026contracts,joy2026add} and calibration \citep{guo2017calibration,angelopoulos2024risk,jia2026balancerag} address related consistency or error targets. Our fixed-score analysis additionally separates individually recoverable decisions from those compatible with one shared threshold.

\section{Conclusion}
Providing a fallback does not ensure reliable handling. Recognized categories can miss their assigned output even when the correct category is supplied; equivalent definitions can redirect decisions while barely changing aggregate scores. PFR tests required output changes and definition consistency on the same inputs, with its four-request pass rate summarizing whether each input satisfies both requirements.

These findings make assignment robustness a concrete target for decision-model evaluation and design. Controlled interventions show why greater agreement or fewer false fallbacks can coexist with lost correct decisions, and why individually recoverable errors may resist a shared selection rule. Reliable fallback use requires verifying the decisions that invoke it.

\section*{Limitations}
This work focuses on identifying and systematically evaluating fallback reliability failures in TDMs rather than developing mitigation methods. Our experimental settings enable systematic comparisons of fallback decisions, but may not capture the full complexity of real-world applications. Future work could explore effective mitigation strategies.
\bibliography{references}

@inproceedings{ribeiro-etal-2020-beyond,
  author={Ribeiro, Marco Tulio and Wu, Tongshuang and Guestrin, Carlos and Singh, Sameer},
  title={Beyond Accuracy: Behavioral Testing of {NLP} Models with {CheckList}},
  booktitle={Proceedings of the 58th Annual Meeting of the Association for Computational Linguistics},
  year={2020},
  pages={4902--4912},
  publisher={Association for Computational Linguistics},
  doi={10.18653/v1/2020.acl-main.442},
  url={https://aclanthology.org/2020.acl-main.442/}
}

@misc{typesafe2026systemone,
  author       = {Almeida, Diogo},
  title        = {Introducing {System One Models} \& {Jev}},
  year         = {2026},
  howpublished = {TypeSafe AI Blog},
  url          = {https://typesafe.ai/blog/introducing-system-one-models-and-jev},
  note         = {Accessed: 2026-10-09}
}

@inproceedings{schick2023toolformer,
title={Toolformer: Language Models Can Teach Themselves to Use Tools},
author={Timo Schick and Jane Dwivedi-Yu and Roberto Dessi and Roberta Raileanu and Maria Lomeli and Eric Hambro and Luke Zettlemoyer and Nicola Cancedda and Thomas Scialom},
booktitle={Thirty-seventh Conference on Neural Information Processing Systems},
year={2023},
url={https://openreview.net/forum?id=Yacmpz84TH}
}

@inproceedings{geng-etal-2023-grammar,
  author={Geng, Saibo and Josifoski, Martin and Peyrard, Maxime and West, Robert},
  title={Grammar-Constrained Decoding for Structured {NLP} Tasks without Finetuning},
  booktitle={Proceedings of the 2023 Conference on Empirical Methods in Natural Language Processing},
  year={2023}, month=dec, pages={10932--10952},
  address={Singapore}, publisher={Association for Computational Linguistics},
  doi={10.18653/v1/2023.emnlp-main.674},
  url={https://aclanthology.org/2023.emnlp-main.674/}
}

@misc{luwu2026coverage,
  author={Lu, Jiawen and Wu, Tongtong},
  title={Benchmarking candidate coverage and rejection policy transfer in typed decision models},
  year={2026}, eprint={2610.03387}, archivePrefix={arXiv},
  note={Version 2}, url={https://arxiv.org/abs/2610.03387v2}
}

@inproceedings{larson-etal-2019-evaluation,
    title = "An Evaluation Dataset for Intent Classification and Out-of-Scope Prediction",
    author = "Larson, Stefan  and
      Mahendran, Anish  and
      Peper, Joseph J.  and
      Clarke, Christopher  and
      Lee, Andrew  and
      Hill, Parker  and
      Kummerfeld, Jonathan K.  and
      Leach, Kevin  and
      Laurenzano, Michael A.  and
      Tang, Lingjia  and
      Mars, Jason",
    editor = "Inui, Kentaro  and
      Jiang, Jing  and
      Ng, Vincent  and
      Wan, Xiaojun",
    booktitle = "Proceedings of the 2019 Conference on Empirical Methods in Natural Language Processing and the 9th International Joint Conference on Natural Language Processing",
    month = nov,
    year = "2019",
    address = "Hong Kong, China",
    publisher = "Association for Computational Linguistics",
    url = "https://aclanthology.org/D19-1131/",
    doi = "10.18653/v1/D19-1131",
    pages = "1311--1316"
}

@inproceedings{casanueva-etal-2020-efficient,
    title = "Efficient Intent Detection with Dual Sentence Encoders",
    author = "Casanueva, I{\~n}igo  and
      Tem{\v{c}}inas, Tadas  and
      Gerz, Daniela  and
      Henderson, Matthew  and
      Vuli{\'c}, Ivan",
    editor = "Wen, Tsung-Hsien  and
      Celikyilmaz, Asli  and
      Yu, Zhou  and
      Papangelis, Alexandros  and
      Eric, Mihail  and
      Kumar, Anuj  and
      Casanueva, I{\~n}igo  and
      Shah, Rushin",
    booktitle = "Proceedings of the 2nd Workshop on Natural Language Processing for Conversational AI",
    month = jul,
    year = "2020",
    address = "Online",
    publisher = "Association for Computational Linguistics",
    url = "https://aclanthology.org/2020.nlp4convai-1.5/",
    doi = "10.18653/v1/2020.nlp4convai-1.5",
    pages = "38--45"
}

@inproceedings{li-roth-2002-learning,
    title = "Learning Question Classifiers",
    author = "Li, Xin  and
      Roth, Dan",
    booktitle = "{COLING} 2002: The 19th International Conference on Computational Linguistics",
    year = "2002",
    url = "https://aclanthology.org/C02-1150/"
}

@misc{zhong2026missing,
  title={When the Right Answer Is Missing: An Arithmetic-Dependent Rejection Bottleneck in Jev},
  author={Zhong, Jike and Li, Ming and Lai, Yuxiang},
  year={2026}, eprint={2609.39496}, archivePrefix={arXiv},
  note={Version 1}, url={https://arxiv.org/abs/2609.39496v1}
}

@misc{chenli2026contracts,
  title={Probability Contracts: Accuracy, Coherence, and Decisions Across LLM Interfaces},
  author={Chen, Han and Li, Yingrui}, year={2026},
  eprint={2609.37470}, archivePrefix={arXiv}, note={Version 2},
  url={https://arxiv.org/abs/2609.37470v2}
}

@misc{joy2026add,
  title={Do System One Decisions Add Up? A Study of Probabilistic Coherence},
  author={Joy, Saman Sarker}, year={2026}, eprint={2609.33971},
  archivePrefix={arXiv}, note={Version 1}, url={https://arxiv.org/abs/2609.33971v1}
}

@inproceedings{pmlr-v139-zhao21c,
  title={Calibrate Before Use: Improving Few-shot Performance of Language Models},
  author={Zhao, Zihao and Wallace, Eric and Feng, Shi and Klein, Dan and Singh, Sameer},
  booktitle={Proceedings of the 38th International Conference on Machine Learning},
  pages={12697--12706}, year={2021}, volume={139},
  series={Proceedings of Machine Learning Research}, publisher={PMLR},
  url={https://proceedings.mlr.press/v139/zhao21c.html}
}

@misc{salinas2024butterfly,
  title={The Butterfly Effect of Altering Prompts: How Small Changes and Jailbreaks Affect Large Language Model Performance},
  author={Salinas, Abel and Morstatter, Fred},
  year={2024}, eprint={2401.03729},
  archivePrefix={arXiv}, note={Version 3},
  url={https://arxiv.org/abs/2401.03729v3}
}

@misc{internlm2026decision,
  author={{InternLM Team}},
  title={{Intern-Decision}: Fast Multi-Modal Decision Model},
  year={2026},
  howpublished={Software and model release},
  url={https://github.com/InternLM/Intern-Decision},
  note={Accessed 9 October 2026}
}

@misc{palmer2026kev4b,
  author={Palmer, Jared},
  title={{Kev-4B}: A Calibrated Decision Model on {Qwen3.5-4B}},
  year={2026},
  howpublished={Hugging Face model card},
  url={https://huggingface.co/jaredpalmer/kev-4b},
  note={Kev 1.0}
}

@misc{theoleecj2026semif,
  author={{TheoLeeCJ}},
  title={{SemIf}: Semantic Ifs from Open Models},
  year={2026},
  howpublished={Software},
  url={https://github.com/TheoLeeCJ/SemIf-OpenJev},
  note={Accessed 9 October 2026}
}

@misc{serikbay2026jevk5,
  author={Serikbay, Alibi},
  title={{JevK5} v0.3: Open-Weight {Jev} Alternative},
  year={2026},
  howpublished={Hugging Face model card},
  url={https://huggingface.co/alibiserikbay/JevK5},
  note={Accessed 9 October 2026}
}

@misc{palmer2026kev08b,
  author={Palmer, Jared},
  title={{Kev-0.8B}: A Calibrated Decision Model on {Qwen3.5-0.8B}},
  year={2026},
  howpublished={Hugging Face model card},
  url={https://huggingface.co/jaredpalmer/kev-0.8b},
  note={Kev 1.0}
}

@misc{convai2026laya,
  author={{Convai Innovations}},
  title={{Laya}},
  year={2026},
  howpublished={Hugging Face model card},
  url={https://huggingface.co/convaiinnovations/laya},
  note={Capacity-screened revision 55cf4c4ebb4e; accessed 9 October 2026}
}

@misc{kotoba2026openjev,
  author={{Kotoba Labs}},
  title={{Open-Jev-DeBERTa-v3-Large}},
  year={2026},
  howpublished={Hugging Face model card},
  url={https://huggingface.co/com-kotobalabs/open-jev-deberta-v3-large},
  note={Capacity-screened revision 188ee67a5c93; accessed 9 October 2026}
}

@inproceedings{guo2017calibration,
  author={Guo, Chuan and Pleiss, Geoff and Sun, Yu and Weinberger, Kilian Q.},
  title={On Calibration of Modern Neural Networks},
  booktitle={Proceedings of the 34th International Conference on Machine Learning},
  series={Proceedings of Machine Learning Research},
  volume={70},
  pages={1321--1330},
  year={2017},
  url={https://proceedings.mlr.press/v70/guo17a.html}
}

@inproceedings{goren2024hierarchical,
  author={Goren, Shani and Galil, Ido and El-Yaniv, Ran},
  title={Hierarchical Selective Classification},
  booktitle={Advances in Neural Information Processing Systems},
  volume={37},
  year={2024},
  url={https://proceedings.neurips.cc/paper_files/paper/2024/hash/c8b100b376a7b338c84801b699935098-Abstract-Conference.html}
}

@misc{li2026llm2jev,
  title={{LLM-as-Jev}: {LLMs} Are Already {Jev}-Style Decision Models---When and How to Fine-Tune Them},
  author={Li, Yinheng and Wagle, Justin},
  year={2026},
  eprint={2610.02076},
  archivePrefix={arXiv},
  primaryClass={cs.CL},
  url={https://arxiv.org/abs/2610.02076v2},
  note={Version 2}
}

@inproceedings{hengst-etal-2024-conformal,
  title={Conformal Intent Classification and Clarification for Fast and Accurate Intent Recognition},
  author={den Hengst, Floris and Wolter, Ralf and Altmeyer, Patrick and Kaygan, Arda},
  booktitle={Findings of the Association for Computational Linguistics: NAACL 2024},
  year={2024},
  pages={2412--2432},
  publisher={Association for Computational Linguistics},
  address={Mexico City, Mexico},
  doi={10.18653/v1/2024.findings-naacl.156},
  url={https://aclanthology.org/2024.findings-naacl.156/}
}

@inproceedings{jain2023amendment,
  title={Test-Time Amendment with a Coarse Classifier for Fine-Grained Classification},
  author={Jain, Kanishk and Karthik, Shyamgopal and Gandhi, Vineet},
  booktitle={Advances in Neural Information Processing Systems},
  volume={36},
  year={2023},
  url={https://proceedings.neurips.cc/paper_files/paper/2023/file/c81690e2cfe63aede8519ad448f56d71-Paper-Conference.pdf}
}

@misc{sun2026typesafe,
  author={Sun, Yu and Xu, Junhao and Shi, Jiajia and Yang, Zijin},
  title={Type-Safe Is Not Error-Free: Typed Decision Models Follow the Option Name, Not the Definition Bound to It},
  year={2026},
  eprint={2609.26758},
  archivePrefix={arXiv},
  primaryClass={cs.AI},
  doi={10.48550/arXiv.2609.26758},
  url={https://arxiv.org/abs/2609.26758v3},
  note={Version 3}
}

@inproceedings{gangrade2021selective,
  author={Gangrade, Aditya and Kag, Anil and Saligrama, Venkatesh},
  title={Selective Classification via One-Sided Prediction},
  booktitle={Proceedings of the 24th International Conference on Artificial Intelligence and Statistics},
  series={Proceedings of Machine Learning Research},
  volume={130},
  pages={2179--2187},
  year={2021},
  publisher={PMLR},
  url={https://proceedings.mlr.press/v130/gangrade21a.html}
}

@inproceedings{pezeshkpour-hruschka-2024-large,
    title = "Large Language Models Sensitivity to The Order of Options in Multiple-Choice Questions",
    author = "Pezeshkpour, Pouya  and
      Hruschka, Estevam",
    editor = "Duh, Kevin  and
      Gomez, Helena  and
      Bethard, Steven",
    booktitle = "Findings of the Association for Computational Linguistics: NAACL 2024",
    month = jun,
    year = "2024",
    address = "Mexico City, Mexico",
    publisher = "Association for Computational Linguistics",
    url = "https://aclanthology.org/2024.findings-naacl.130/",
    doi = "10.18653/v1/2024.findings-naacl.130",
    pages = "2006--2017"
}

@inproceedings{wei-etal-2024-unveiling,
    title = "Unveiling Selection Biases: Exploring Order and Token Sensitivity in Large Language Models",
    author = "Wei, Sheng-Lun  and
      Wu, Cheng-Kuang  and
      Huang, Hen-Hsen  and
      Chen, Hsin-Hsi",
    editor = "Ku, Lun-Wei  and
      Martins, Andre  and
      Srikumar, Vivek",
    booktitle = "Findings of the Association for Computational Linguistics: ACL 2024",
    month = aug,
    year = "2024",
    address = "Bangkok, Thailand",
    publisher = "Association for Computational Linguistics",
    url = "https://aclanthology.org/2024.findings-acl.333/",
    doi = "10.18653/v1/2024.findings-acl.333",
    pages = "5598--5621"
}

@article{
chen2024frugalgpt,
title={Frugal{GPT}: How to Use Large Language Models While Reducing Cost and Improving Performance},
author={Lingjiao Chen and Matei Zaharia and James Zou},
journal={Transactions on Machine Learning Research},
issn={2835-8856},
year={2024},
url={https://openreview.net/forum?id=cSimKw5p6R},
note={Featured Certification}
}

@inproceedings{
ong2025routellm,
title={Route{LLM}: Learning to Route {LLM}s from Preference Data},
author={Isaac Ong and Amjad Almahairi and Vincent Wu and Wei-Lin Chiang and Tianhao Wu and Joseph E. Gonzalez and M Waleed Kadous and Ion Stoica},
booktitle={The Thirteenth International Conference on Learning Representations},
year={2025},
url={https://openreview.net/forum?id=8sSqNntaMr}
}

@inproceedings{mozannar2020defer,
  title={Consistent Estimators for Learning to Defer to an Expert},
  author={Mozannar, Hussein and Sontag, David},
  booktitle={Proceedings of the 37th International Conference on Machine Learning},
  series={Proceedings of Machine Learning Research},
  volume={119},
  pages={7076--7087},
  year={2020},
  publisher={PMLR},
  url={https://proceedings.mlr.press/v119/mozannar20b.html}
}

@inproceedings{saeidi2018sharc,
  title={Interpretation of Natural Language Rules in Conversational Machine Reading},
  author={Saeidi, Marzieh and Bartolo, Max and Lewis, Patrick and Singh, Sameer and Rockt{\"a}schel, Tim and Sheldon, Mike and Bouchard, Guillaume and Riedel, Sebastian},
  booktitle={Proceedings of the 2018 Conference on Empirical Methods in Natural Language Processing},
  pages={2087--2097},
  year={2018},
  publisher={Association for Computational Linguistics},
  url={https://aclanthology.org/D18-1233/},
  doi={10.18653/v1/D18-1233}
}

@inproceedings{angelopoulos2024risk,
  title={Conformal Risk Control},
  author={Angelopoulos, Anastasios N. and Bates, Stephen and Fisch, Adam and Lei, Lihua and Schuster, Tal},
  booktitle={International Conference on Learning Representations},
  year={2024},
  url={https://openreview.net/forum?id=33XGfHLtZg}
}

@misc{yu2026salmonnduo,
  title={{SALMONN-duo}: Adaptive Dual-System Coordination for Full-Duplex Voice Agents},
  author={Yu, Wenyi and Wang, Siyin and Chiba, Terumi and Chen, Xianzhao and Tian, Xiaohai and Zhang, Jun and Lu, Lu and Zhang, Chao},
  year={2026},
  eprint={2609.34247},
  archivePrefix={arXiv},
  note={Version 1},
  url={https://arxiv.org/abs/2609.34247}
}

@inproceedings{geifman2019selectivenet,
  author={Geifman, Yonatan and El-Yaniv, Ran},
  title={{SelectiveNet}: A Deep Neural Network with an Integrated Reject Option},
  booktitle={Proceedings of the 36th International Conference on Machine Learning},
  series={Proceedings of Machine Learning Research},
  volume={97},
  pages={2151--2159},
  year={2019},
  publisher={PMLR},
  url={https://proceedings.mlr.press/v97/geifman19a.html}
}

@misc{huang2026rlcascade,
  author={Huang, Shihong and Wang, Shengjie and Ma, Hong and Xu, Zhou},
  title={{RLCascadeRouter}: Quality-Estimator-Free Cascade Routing via Reinforcement Learning},
  year={2026},
  eprint={2608.15817},
  archivePrefix={arXiv},
  primaryClass={cs.AI},
  doi={10.48550/arXiv.2608.15817},
  url={https://arxiv.org/abs/2608.15817v1}
}

@misc{jia2026balancerag,
  author={Jia, Zijun and Ye, Yuanchang and Jia, Sen and Qian, Yiyao and Wang, Haoning and Chen, Baojie and Tang, Diyin and Yu, Jinsong and Wang, Zhiyuan},
  title={{BalanceRAG}: Joint Risk Calibration for Cascaded Retrieval-Augmented Generation},
  year={2026},
  eprint={2605.20084},
  archivePrefix={arXiv},
  primaryClass={cs.CL},
  doi={10.48550/arXiv.2605.20084},
  url={https://arxiv.org/abs/2605.20084v1}
}

@misc{hu2026jevadvbench,
      title={JevAdvBench: A Benchmark and Black-Box Attacks for Reinforcement Learning for Calibrated Decisions Models}, 
      author={Jianyi Hu and Hangtao Zhang and Yi Liu and Yeqi Zeng and Li Zeng and Xianlong Wang and Rui Wang and Leo Yu Zhang},
      year={2026},
      eprint={2609.31142},
      archivePrefix={arXiv},
      primaryClass={cs.CR},
      url={https://arxiv.org/abs/2609.31142}, 
}

@misc{wu2026decisionhijacking,
      title={Decision Hijacking: Prompt Injection Attacks on Jev's Typed Probabilistic Decisions}, 
      author={Tiantong Wu and Wei Yang Bryan Lim},
      year={2026},
      eprint={2609.28613},
      archivePrefix={arXiv},
      primaryClass={cs.CR},
      url={https://arxiv.org/abs/2609.28613}, 
}
\appendix
\section{Evaluation Protocol}

\begin{figure*}[t]
\centering
\includegraphics[width=\textwidth]{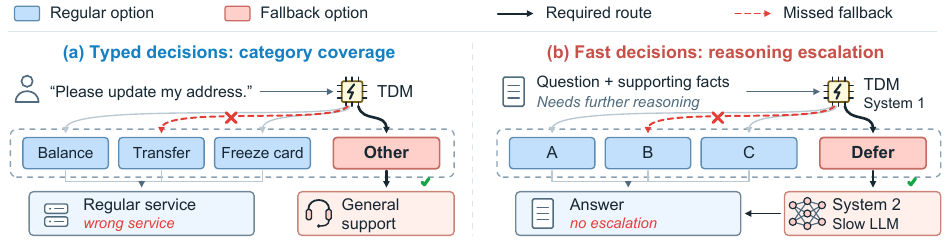}
\caption{A TDM serves as (a) a category-routing model or (b) a fast System 1. Fallback options control access to general support or System 2; missed fallbacks bypass the intended handling.}
\label{fig:fallback-use-cases}
\end{figure*}

\label{app:setup}\label{app:models}\label{app:accounting}\label{app:criteria}\label{app:menus}\label{app:request-construction}
\paragraph{Sampling and pairing.}
Fixed hash ranks select three CLINC and five BANKING test texts per class; CLINC out-of-scope inputs are excluded. TREC's test split covers 42 of the 50 inventory categories. Diagnostic texts come from CLINC validation and BANKING/TREC training, using within-category ranks and round-robin selection. Exact and normalized matches to test, development, calibration and other diagnostic inputs are excluded. Development and calibration pools instead preserve the eligible source distribution (Appendix~\ref{app:risk-calibration}).

Category-information/routing, wording, matched-length, choice-set/position and naming comparisons share one diagnostic panel per dataset. Replacement/layout, policy, boundary and cross-backbone panels use disjoint texts. Reusing inputs within a comparison holds the category distribution fixed while changing the tested factor. Fallback positions cycle over the sixteen slots and remain fixed within each pair, except in the explicit position control.

\paragraph{Category descriptions.}
CLINC and BANKING use the released intent names with underscores rendered as spaces. TREC uses the dataset authors' fine-category definitions, aligning \texttt{creative} to the released \texttt{ENTY:cremat} identifier. Requests supply the full inventory; category and union-member lists follow lexicographic order. Fallback criteria appear in Table~\ref{tab:wording-criteria}.

\paragraph{Model interfaces.}
Intern scores decision symbols \citep{internlm2026decision}, Kev uses an option-pointer head \citep{palmer2026kev4b,palmer2026kev08b}, and JevK5 and SemIf score answer letters \citep{serikbay2026jevk5,theoleecj2026semif}. Native option limits are 62 for Intern and 255 for Kev. Inference uses bfloat16 and score temperatures of 1 (Intern/SemIf), 1.22 (JevK5) and 2.40605 (Kev). Model parameters remain fixed; threshold analyses reuse the evaluation scores.

JevK5 uses v0.3; Kev is built on Qwen3.5-4B-Base, while SemIf uses Qwen3.5-4B.

\paragraph{Statistical units.}
\label{app:statistics}
Tests and resampling retain each text's variants together. Primary U/C contrasts use two-sided exact McNemar tests on $A_-$ and $F_+$. Holm correction covers all 42 tests: two metrics, three datasets and seven parameter configurations (the four 4B settings, Intern 0.8B/2B and Kev 0.8B). Diagnostic intervals use 10,000 paired text-bootstrap draws, averaging mappings, templates or set/position combinations within texts. Compact-task resampling preserves coarse-category quotas. One repeat of the 384 matched natural-input requests per 4B configuration produced identical choices.

\section{Recognition and Model Scope}
\label{app:scale}\label{app:known-protocol}\label{app:recognition}
\paragraph{Category information.}
The three input conditions retain the text, add its correct category, or provide the category alone (Figure~\ref{fig:information}). They distinguish ordinary recognition from selecting the assigned output once the category is supplied. Figure~\ref{fig:information-full} gives all three datasets.

\paragraph{Paired examples.}
\label{app:paired-examples}
Intern correctly identifies both inputs below in the full-inventory reference and under both gold-regular definitions. For ``How far is it from Denver to Aspen?'', replacing \texttt{NUM:dist} with \texttt{LOC:mount} makes fallback correct: U selects fallback, but C selects \texttt{LOC:mount}. For ``What is home equity?'', replacing \texttt{DESC:def} with \texttt{NUM:weight} instead yields \texttt{ENTY:veh} under U and fallback under C.

\paragraph{Parameter counts and backbones.}
Intern 0.8B/2B/4B and Kev 0.8B/4B use identical main evaluation texts and assignments (Figure~\ref{fig:scale-extension}). Each point represents a separately trained model with its native readout and calibration. The matched cross-backbone task uses 256 TREC texts: 43 each from ABBR, DESC, ENTY and HUM, and 42 each from LOC and NUM. It retains six categories, four options, matched criteria and balanced positions; all inputs fit each model's input limit.

\begin{figure*}[!t]
\centering
\includegraphics[width=\textwidth]{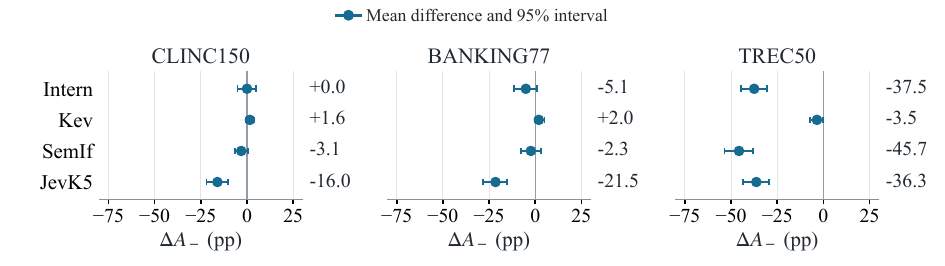}
\caption{Definition sensitivity survives length matching. Negative-minus-positive fallback recall (pp), averaged over both T1/T2 mappings; bars show pointwise paired-text intervals.}
\label{fig:matched}
\end{figure*}

\section{Definition Robustness Controls}
\label{app:robustness}\label{app:sensitivity}\label{app:complete}\label{app:matched}\label{app:matched-criteria}\label{app:surface-controls}\label{app:semantic-layout}
Table~\ref{tab:wording-criteria} lists the three criterion pairs. Pair 1 defines the main evaluation. Requests specify that each input belongs to exactly one inventory category and that fallback denotes a category group; regular criteria give the category label and definition.

\begin{table*}[t]
\centering
\begin{tabular}{@{}c p{.435\textwidth} p{.435\textwidth}@{}}
\toprule
Pair & Union (U) & Complement (C)\\
\midrule
1 & the input belongs to any one of these categories in U: [members]. & the input belongs to a category in U that is not represented by any of the other output options.\\
\addlinespace[3pt]
2 & choose this option when the input belongs to one of the following categories in U: [members]. & choose this option when the input's category in U is outside the categories represented by the other output options.\\
\addlinespace[3pt]
3 & this option covers exactly these categories in U: [members]. & this option covers exactly the categories in U other than those represented by the other output options.\\
\bottomrule
\end{tabular}
\caption{Three pairs of logically equivalent fallback definitions. In the quoted criteria, $\univ$ denotes the full category inventory, not the Union formulation. The common fallback-option name is omitted; [members] lists fallback categories. All other request fields remain fixed.}
\label{tab:wording-criteria}
\end{table*}

\paragraph{Length and category mentions.}
A shared membership table partitions the inventory into T1/T2. Positive wording selects inputs in the designated group; negative wording excludes the other group. Swapping the group identifiers gives a second comparison. All 6,144 positive/negative request pairs match token counts and category mentions. Figure~\ref{fig:matched} shows that the resulting recall differences depend on both model and dataset.

\begin{figure*}[!t]
\centering
\includegraphics[width=\textwidth]{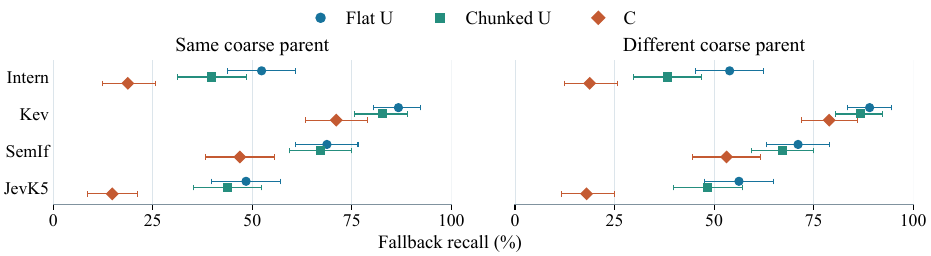}
\caption{Replacement categories and list layout affect fallback recall on TREC. Flat and chunked U preserve membership and order; same/different parent refers to the gold category's replacement. C is unchanged by chunking. Whiskers show paired-text intervals.}
\label{fig:semantic-layout}
\end{figure*}

\paragraph{Replacement and list layout.}
On TREC's full inventory, the gold option is replaced in the same slot by a category from either the same or a different coarse parent. The other fourteen regular options stay fixed. In the layout comparison, six spaces in the 35-member union list become line breaks, preserving membership, order and punctuation. For Intern with different-parent replacement, this change repairs one decision but harms 21, lowering recall from 53.9\% to 38.3\% (Figure~\ref{fig:semantic-layout}).

\begin{figure}[t]
\centering
\includegraphics[width=\columnwidth]{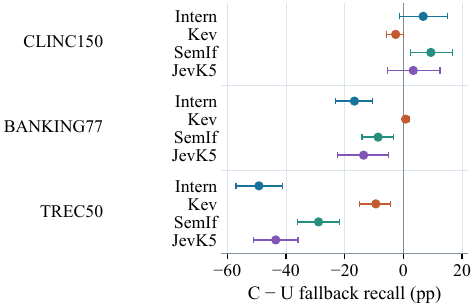}
\caption{Definition effects across three wording pairs. C-minus-U fallback recall averages templates within texts; bars show paired-text intervals.}
\label{fig:wording}
\end{figure}

\paragraph{Wording, choice sets and position.}
Three U/C wording pairs are crossed with both assignments (Figure~\ref{fig:wording}). A separate comparison crosses three choice sets with four fallback positions. C's TREC recall loss persists in all twelve settings for every configuration, averaging 10.0--46.4 points. 
Across three choice sets, changing the fallback position alters 70.8\% of SemIf/CLINC gold-fallback decisions under U, compared with 10.9\% under C. Figure~\ref{fig:position-interaction} illustrates a separate position effect for JevK5/CLINC, where the U/C recall gap reverses across positions.

\paragraph{Fallback names.}
\label{app:fallback-names}
Seven fallback names are compared with the complement criterion, category membership and positions fixed (Figure~\ref{fig:fallback-names}). Effects vary across configurations: None lowers SemIf/TREC recall from 53.1\% to 29.7\%, whereas None of the above improves Intern/TREC recall. For SemIf/TREC, Other and None respectively yield 13 and 8 false fallbacks, but correct assignment pairs fall from 56 to 29 out of 128 ($J$: 43.8\% to 22.7\%).

\paragraph{Individual decision changes.}
Figure~\ref{fig:transitions} decomposes U/C disagreements into correct-to-wrong, wrong-to-correct, and wrong-to-wrong changes. Opposing changes can offset each other in aggregate accuracy, concealing substantial differences in individual decisions.

\section{Policy Execution}
\subsection{Policies from Quality and Cost}
\label{app:defer-policy}
TREC texts are assigned to twelve parameter combinations, with 10 or 11 texts each: $q_{\mathrm{fast}}\in\{0.50,0.65,0.80\}$, gain $d\in\{0.10,0.15\}$ and cost $c\in\{0.10,0.20\}$. We set $q_{\mathrm{slow}}=q_{\mathrm{fast}}+d$ and compare $\lambda=d/c+0.25$ (Fast required) with $\lambda=d/c-0.25$ (Defer required). Text, estimates and action order stay fixed. Gain wording selects Defer when $d>\lambda c$; utility-complement wording selects Defer unless $q_{\mathrm{fast}}\geq q_{\mathrm{slow}}-\lambda c$. Both select Fast on ties.

\begin{table}[t]
\centering
\begin{tabular*}{\columnwidth}{@{\extracolsep{\fill}}l@{\,}cc@{\;}cc@{\;}cc@{}}
\hline
\multirow{2}{*}{Model} & \multicolumn{2}{c}{Fast correct} & \multicolumn{2}{c}{Defer correct} & \multicolumn{2}{c}{DR}\\
\cline{2-3}\cline{4-5}\cline{6-7}
& Gain & Util. & Gain & Util. & Fast & Defer\\
\hline
Intern & 125 & 68 & 3 & 29 & 57 & 26\\
Kev & 105 & 109 & 14 & 9 & 12 & 11\\
SemIf & 128 & 128 & 0 & 0 & 0 & 0\\
JevK5 & 128 & 128 & 0 & 0 & 0 & 0\\
\hline
\end{tabular*}
\caption{Policy decisions from supplied estimates, counts out of 128 texts. Gain/Util. denote equivalent gain and utility-complement criteria. Correct counts are separated by the required action; DR counts disagreements between criteria under Fast-required or Defer-required policies. Both policy $J$ and all-four correctness are zero throughout.}
\label{tab:defer-parent}
\end{table}

Each input is evaluated with estimates alone, computed Fast/Defer utilities, or the correct Defer predicate and action mapping. The two policy weights require Fast and Defer, respectively, and joint correctness $J$ requires both decisions to be correct (Figure~\ref{fig:routing}). Equal weighting of the twelve parameter combinations preserves these findings.

For ``How many James Bond novels are there?'', Kev receives $q_{\mathrm{fast}}=0.80$, $q_{\mathrm{slow}}=0.95$, $c=0.20$ and $\lambda=0.50$, making Defer optimal ($0.85>0.80$). Under gain wording, estimates alone elicit Fast, computed utilities elicit Defer, and the correct predicate elicits Fast.

\paragraph{Boundary conditions.}
Separate panels use 64 texts each for zero gain, negative gain and utility ties. Intern selects Fast correctly under all four requests on 32, 61 and 38 texts, respectively, versus 62--64 for the other models. The zero-gain example in Section~\ref{sec:routing} shows that errors can also introduce unnecessary deferral.

\begin{figure*}[p]
\centering
\includegraphics[width=\textwidth]{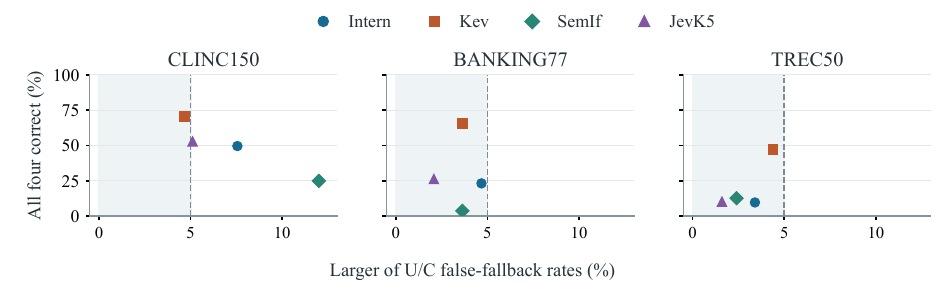}
\caption{Low false fallback does not ensure PFR correctness. Each point is a model--dataset setting under the shared calibration threshold. Shading marks test $F_+\leq5\%$ under both U/C; the vertical axis requires correctness on all four requests.}
\label{fig:risk-pfr}

\par
\includegraphics[width=\textwidth]{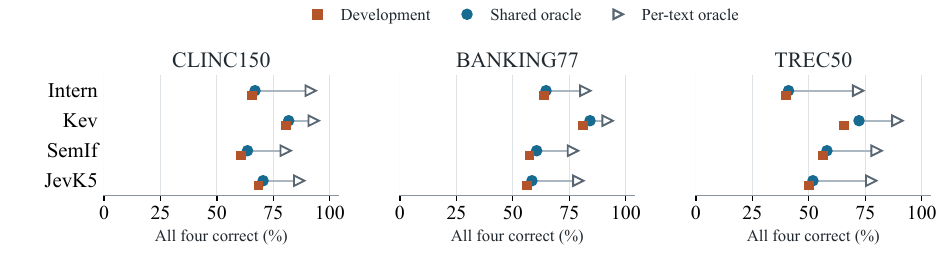}
\caption{Individual recoverability exceeds shared-threshold recovery in every setting. All-four correctness under the development-selected threshold, best shared threshold, or a separate threshold per text. The latter two use test labels and bound recovery at fixed scores. Error decomposition: Figure~\ref{fig:boundary-partition}.}
\label{fig:recovery-limits}

\par\medskip
\begin{minipage}[t]{\columnwidth}
\centering
\includegraphics[width=\columnwidth]{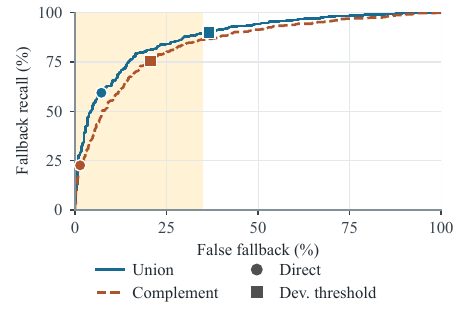}
\caption{Intern/TREC recall--false-fallback trade-off at fixed scores. Marks indicate direct selection and the development threshold. Shading highlights 0--35\% false fallback.}
\label{fig:threshold-example}
\end{minipage}\hfill
\begin{minipage}[t]{\columnwidth}
\centering
\includegraphics[width=\columnwidth]{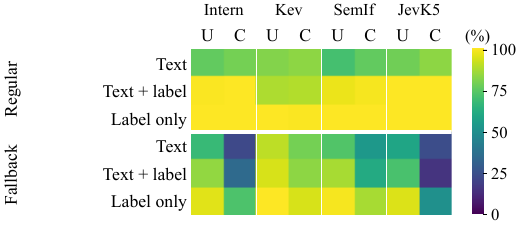}
\caption{Category information on TREC. Each model has U/C columns. Text+label adds the gold category; Label only removes the text. Full comparison: Figure~\ref{fig:information-full}.}
\label{fig:information}
\end{minipage}
\end{figure*}

\begin{figure*}[!t]
\centering
\includegraphics[width=\textwidth]{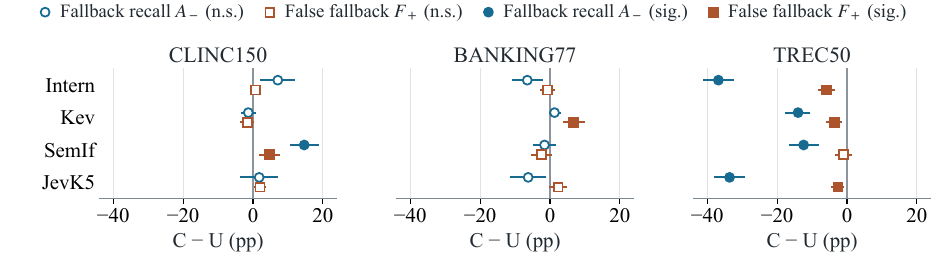}
\caption{Aggregate U/C differences in the main evaluation. Complement-minus-union score changes with paired-text intervals. Filled markers survive the 42-test Holm correction.}
\label{fig:primary}
\par\medskip
\centering
\includegraphics[width=\textwidth]{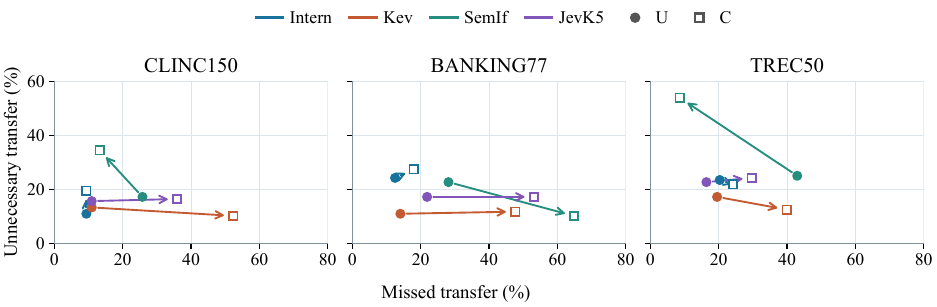}
\caption{Equivalent category policies change missed and unnecessary transfers. Arrows connect U (filled circle) to C (open square); lower-left is better. Missed transfer is $1-A_-$ on gold-System2, and unnecessary transfer is $F_+$ on gold-Fast.}
\label{fig:route-errors}
\end{figure*}

\subsection{Policies from Category Membership}
\label{app:routing-protocol}
A randomly selected half of the category inventory maps to Fast and the remainder to System2. Swapping gold with one category in the other group changes the required action while preserving group sizes and action order. U lists System2 members; C excludes Fast members. The input's category is not disclosed. These binary choices imply $A_++F_+=1$. Gold-System2 disagreement spans 4.7\%--44.5\% across the twelve model--dataset pairs (Figure~\ref{fig:route-errors}).

\section{Selection Rules and Recovery}
\subsection{Fixed Scores and Shared Thresholds}
\label{app:recovery-limits}\label{app:rank-actions}\label{app:threshold}\label{app:direct-threshold}\label{app:score-details}
\begin{figure*}[!t]
\centering
\centering
\includegraphics[width=\textwidth]{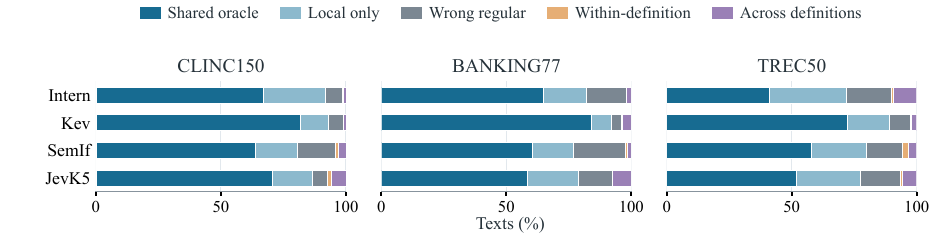}
\caption{Fixed-score recovery has local and shared-rule limits. Bars partition the main evaluation texts: maximal shared-threshold coverage; additional individually recoverable cases; then failures assigned to wrong regular winners, within-definition conflict or cross-definition conflict.}
\label{fig:boundary-partition}

\par\medskip
\includegraphics[width=\textwidth]{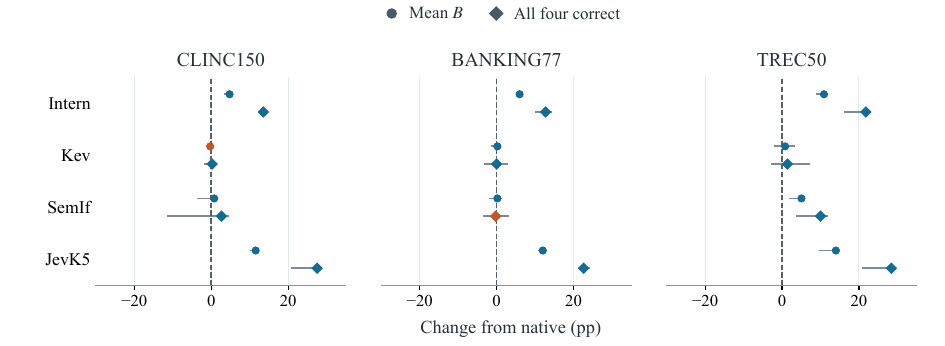}
\caption{Recovery varies with the development sample. Circles (mean $B$) and diamonds (all-four correctness) show median changes from direct selection; lines give empirical 2.5th--97.5th percentiles over development-pool refits. Blue/orange denotes improvement/deterioration.}
\label{fig:dev-sensitivity}

\par\medskip
\centering
\includegraphics[width=\textwidth]{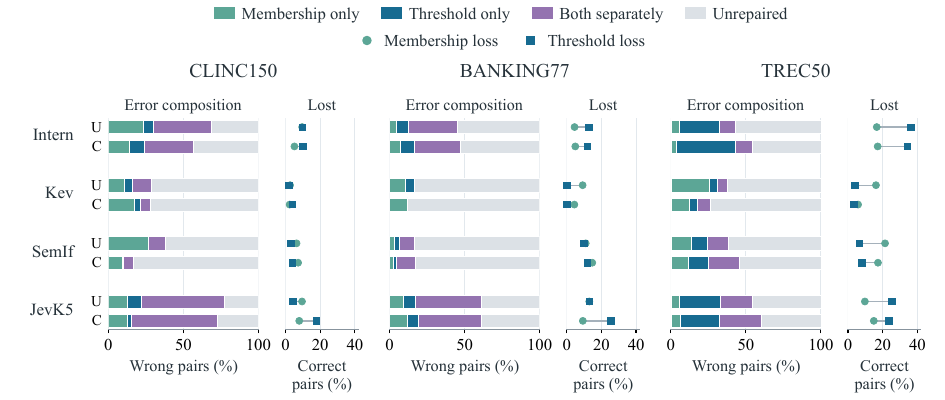}
\caption{Interventions both repair and harm decisions. Left bars partition initially wrong pairs; right points give the percentage of initially correct pairs overturned. Aligned rows share model and definition, but use different denominators. Both means a pair repaired separately by each intervention.}
\label{fig:repair-overlap}
\end{figure*}

Let $m_e^+$/$m_e^-$ be the fallback-minus-best-regular log-score margin under gold-regular/fallback for $e\in\{U,C\}$. A definition's pair is correct iff its gold-regular request has the correct regular winner and $\tau\in[m_e^+,m_e^-)$. All four requests are correct iff the gold-regular winners under U and C are both correct and
\begin{equation}
\tau\in[\max(m_U^+,m_C^+),\min(m_U^-,m_C^-)).
\end{equation}
For texts satisfying this regular-winner condition, a nonempty interval makes all four requests individually recoverable. The largest overlap among these eligible intervals gives the best shared threshold. These bounds optimize four-request correctness, whereas the development rule optimizes mean $B$ (Figure~\ref{fig:recovery-limits}). Figure~\ref{fig:boundary-partition} partitions failures into incorrect regular winners and incompatible margins within or across definitions.

\subsection{One-Sided Calibration}
\label{app:risk-calibration}
Calibration and balanced-threshold fitting use the same development panels, disjoint from all evaluation texts. Sampling preserves the eligible source distribution; nested subsets of 32 and 64 assess development-size sensitivity. For gold-regular margins $m=\log p_{\mathrm{fallback}}-\log\max p_{\mathrm{regular}}$, a threshold selects fallback when $m>\tau$. At level $\alpha$, $\tau$ is the $\lceil(n+1)(1-\alpha)\rceil$th order statistic, or $+\infty$ when its rank exceeds $n$. The rule controls false fallback under exchangeability of calibration and evaluation scores \citep{angelopoulos2024risk}. We fit one threshold per definition and use their maximum; under exchangeability for each definition, this retains both marginal one-sided bounds.

The shared threshold matters: Kev/BANKING's U-only threshold leaves C false fallback above the 5\% budget, whereas the maximum brings both below it. Across datasets, the calibration and target populations change: 19 of 48 target-definition comparisons exceed the empirical budget at $\alpha=5\%$, reaching 17.4\%. Transfer comparisons share target texts.

\subsection{Development Sensitivity and Recovery Costs}
\label{app:dev-sensitivity}\label{app:recovery-costs}
Refitting thresholds on 2,000 paired bootstrap samples keeps each text's four requests together and holds test scores fixed. The median refit improves mean $B$ in eleven settings; seven improve it in at least 95\% of refits (Figure~\ref{fig:dev-sensitivity}). For Intern/TREC, every refit improves $B$ and gold-fallback agreement, while increasing gold-regular disagreement.

\paragraph{Threshold fitting.}
Candidate thresholds are development-margin midpoints, zero and extreme endpoints; probabilities are floored at $10^{-12}$. Ties prefer fewer false fallbacks, smaller threshold magnitude, then the larger threshold. Choice order breaks regular-score ties. On the 21,360 main requests, each native choice is a maximum-probability option. First-in-order argmax agrees on 21,333 requests; the strict zero-threshold rule agrees on 21,263. All differences involve tied maxima. For Intern/TREC, complement recall and false fallback are unchanged at zero threshold (22.6\% and 1.4\%).

\begin{figure*}[p]
\centering
\includegraphics[width=\textwidth]{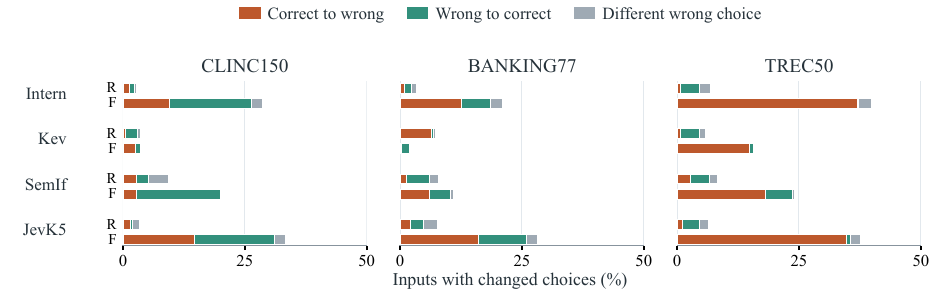}
\caption{Score differences conceal opposing transitions. Bars partition U-to-C choice changes as percentages of all texts under each gold assignment (R: regular; F: fallback). Different-wrong excludes unchanged wrong choices. Segments sum to DR; repairs and harms can cancel in accuracy.}
\label{fig:transitions}

\par\medskip
\centering
\includegraphics[width=\textwidth]{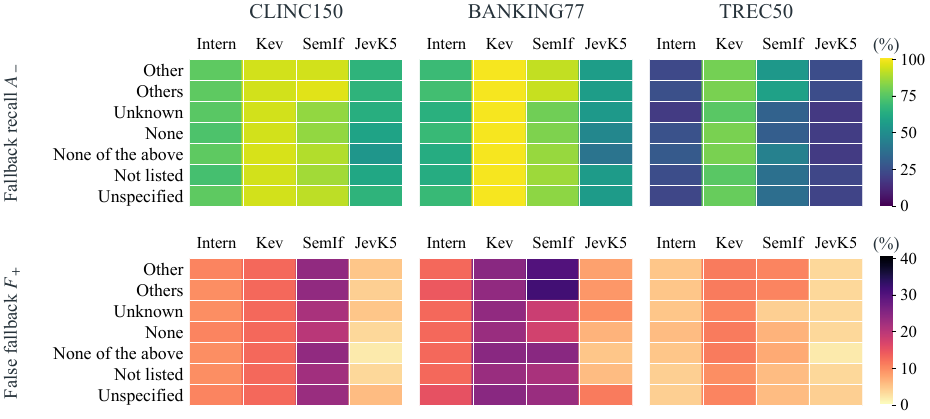}
\caption{Fallback-name effects with the complement criterion fixed. Inputs, membership and positions are paired across seven names. Top: fallback recall $A_-$. Bottom: false fallback $F_+$. Both are percentages; their color scales differ.}
\label{fig:fallback-names}

\par\medskip
\includegraphics[width=\textwidth]{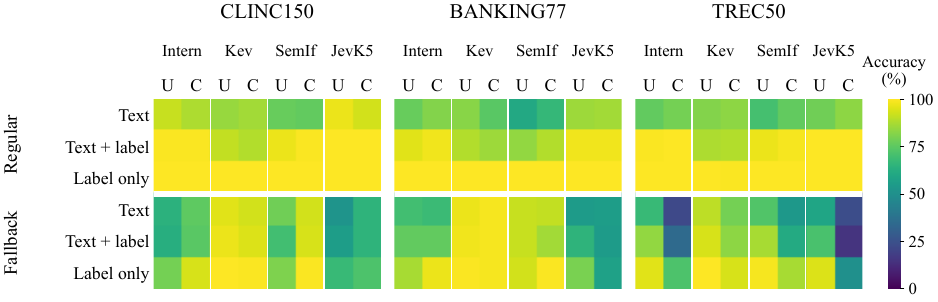}
\caption{Category information across all datasets, extending Figure~\ref{fig:information}. Columns pair U/C within each model; rows show the three input modes under regular or fallback assignment. Colors use a shared accuracy scale.}
\label{fig:information-full}
\end{figure*}

\end{document}